\documentclass{appolb}
\usepackage{epsfig}
\usepackage[english]{babel}
\usepackage[utf8]{inputenc}
\usepackage{amssymb}

\usepackage{color}

\newcommand{\ak}{a^\dagger}

\newcommand{\ra}{\rangle}
\newcommand{\la}{\langle}

\newcommand{\beq}{\begin{eqnarray}}
\newcommand{\eeq}{\end{eqnarray}}
\newcommand{\nn}{\nonumber}
\newcommand{\es}{& = &}

\begin{document}

\title{Application of a numerical renormalization group 
	   procedure to an elementary anharmonic oscillator
\thanks{Based on the author's bachelor thesis, 
        University of Warsaw, July 2010.}
\author{K. P. W\'ojcik}
\address{ Faculty of Physics, Adam Mickiewicz University, 
	      61-614 Pozna\'{n}, Poland }
}
\date{\today}
\headtitle{Numerical RG and anharmonic oscillator}
\headauthor{K.~P.~Wójcik}

\maketitle

\begin{abstract}

The canonical quantum Hamiltonian eigenvalue problem for 
an anharmonic oscillator with a Lagrangian $L = \dot 
\phi^2/2 - m^2 \phi^2/2 - g m^3 \phi^4$ is numerically solved 
in two ways. One of the ways uses a plain cutoff on the 
number of basis states and the other employs a renormalization 
group procedure. The latter yields superior results to the 
former because it allows one to calculate the effective 
Hamiltonians. Matrices of effective Hamiltonians are quite small in comparison 
to the initial cutoff but nevertheless yield accurate 
eigenvalues thanks to the fact that just eight of their 
highest-energy matrix elements are proper functions of 
the small effective cutoff. We explain how these cutoff-dependent 
matrix elements emerge from the structure of the Hamiltonian
and the renormalization group recursion, and we show that 
such small number of cutoff-dependent terms is sufficient
to renormalize any band-diagonal Hamiltonian.

\end{abstract}

\section{Introduction}

The purpose of this article is to show that a renormalization group procedure 
(RGP) of Hamiltonians, based on Gaussian elimination of basis states~\cite{Wilson70},
provides a convenient approach to the Hamiltonian matrix diagonalization problem 
for an elementary oscillator with quartic anharmonicity, 
\beq
H \es {\dot \phi^2 \over 2} + {m^2 \phi^2 \over 2} + g m^3 \phi^4,
\label{H}
\eeq
where $g>0$ is a dimensionless coupling constant and the units are chosen such 
that $\hslash=1$ and $c=1$. By ``elementary'' we mean that we consider $\phi$ 
a single variable, describing only one mode of a scalar field. Thus, instead of 
a quantum field theory problem, we have a simple quantum mechanics problem. 
The utility of RGP results from the band-diagonal structure of the oscillator's 
Hamiltonian. Having written the initial Hamiltonian in the second-quantized form,
\beq
H = m \left[ \ak a + g(\ak+a)^4 \right]
\label{Ha}
\eeq
(the irrelevant constant $+m/2$ was suppressed), the Hamiltonian matrix is 
obtained by evaluating matrix elements in the basis of normalized 
eigenstates of $H_0 = m \ak a$. 

In general, band-diagonal structure of a Hamiltonian corresponds to the 
situation, when an interaction cannot mix the states belonging to distinct 
energy scales. Such an interaction is much easier to understand then the 
one which does not possess this property. For this reason, building a
unitary transformation which decouples states distinct in energy from each 
other may be the main part of solving a Hamiltonian eigenvalue problem 
involving interactions of the latter kind. The examples of methods dealing 
with such unitary decoupling are the similarity renormalization group (SRG) 
scheme \cite{SRG,RG+SRG} and (closely related) Wegner's flow equation 
\cite{Wegner,Kehrein,Furnstahl}. Both methods bring (arbitrarily chosen) 
initial Hamiltonian to a band-diagonal form. The oscillator Hamiltonian possesses
a band diagonal structure from the beginning, i.e., without any need for using
SRG to make it so. Thus, this paper can be considered as dealing with the
second stage of solving the eigenvalue problem after the band diagonal 
structure is already achieved.

Namely, in this paper we address the problem of precise and systematic numerical 
calculation of small eigenvalues of the band-diagonal Hamiltonian, using
the one in Eq.~(\ref{Ha}) as an example. We show that the Wilsonian RGP in 
the case of any band-diagonal Hamiltonian is reduced to a recursive relation 
for only a few (six in the considered example) independent Hamiltonian matrix 
elements. Thus, what we do here may be considered a convenient tool for 
analysing band-diagonal Hamiltonians emerging from aforementioned methods. 
Further discussion of this possibility in the case of realistic field-theoretic Hamiltonians can be found in~\cite{RG+SRG}, see Appendix~A there.

In zero-dimensional oscillator there are no ultraviolet divergences, 
so the renormalization issues in this model are limited to finite cutoff 
dependence. Such finite cutoff dependence strictly disappears from the 
smallest eigenvalues when the cutoff is sent to infinity. Thus, there is 
no need to introduce diverging counterterms. Nevertheless, RGP is still 
of great value, because it {\it gets rid of the spurious finite cutoff 
dependence of small Hamiltonian eigenvalues}, even for very small cutoffs. 
The numerical recipe is very fast, precise and nonperturbative. So, it is 
valid for all values of $g$. The low-energy spectra of the Hamiltonian~(\ref{Ha}) 
obtained with RGP method will be compared with the corresponding spectra 
obtained by simply cutting off the initial Hamiltonian at the same small 
cutoff, for different values of $g$.

RGPs are extensively used as numerical tools since Wilson has delivered the 
algorithm \cite{Wilson75} for treating the systems in which Kondo effect
\cite{Kondo} appears. The algorithm considered in this paper resembles the 
one considered in \cite{Wilson75}, but does not include a logarithmic discretization 
of any conduction band. It does, however, deal with a discrete set of states
that is similar to an infinite set of intervals of pion momentum in~\cite{Wilson65} 
and a discrete set in \cite{Wilson70}. 
Ref.~\cite{survey} provides an overview of applications of numerical RGPs 
to the quantum impurity systems. Here we show that an appropriate numerical 
RGP can be always applied to any quantum Hamiltonian that is band-diagonal. 
In particular, we have in mind systems of particles for which SRG renders 
a Hamiltonian of the band-diagonal form \cite{RG+SRG}.

Our approach is not the only one possible in the case of a quartic oscillator. 
Another approach, based directly on SRG, was presented in~\cite{RJP}. In the 
case of oscillator, the procedure of Ref.~\cite{RJP} is simple and 
an effective generator of SRG transformation 
is found using specific simplifications. Most of the work is done analytically. 
However, for an  arbitrary band-diagonal Hamiltonian such simple generator 
does not exist and numerical calculations based on SRG are complex. In fact, 
in such circumstances they are less effective then these based on Wilsonian 
RGP that are presented in this article using the example of quartic oscillator.

The paper is organized as follows. The RGP carried out in this article is 
described in Section~\ref{sec:RG}. The resulting evolution of Hamiltonian 
matrix elements is described in Section~\ref{sec:course}. Section~\ref{comparison} 
compares results obtained using RGP with results obtained using a plain 
cutoff (PC) of the Hamiltonian matrix, without inclusion of any corresponding 
changes in its matrix elements. The PC procedure has no a priori justification
for general Hamiltonians but works for large enough cutoffs for band-diagonal 
Hamiltonians, see Section~\ref{comparison} for details. The effective 
Hamiltonians we obtain are described in Section~\ref{EH}. Section~\ref{Conclusion} 
concludes the article with a summary of accuracy achieved using the RGP.

\section{ Description of RGP carried out in this article }
\label{sec:RG}

We set $m=1$ and denote $H_I= (H-H_0)/g$. In the basis of normalized 
eigenstates of $H_0$, \hbox{$|k\ra = (k!)^{-1/2}(\ak)^k |0\ra$}, $H$ 
has matrix elements
\beq
\label{Hmn}
H_{kl} \equiv \la k | H | l \ra
	\es [k + 3g(2k^2+2k+1)] \delta_{kl} +  	\\
		&&+\,\, g(4l+6) \sqrt{(l+1)(l+2)} \delta_{k (l+2)} + \nn \\
		&&+\,\, g(4k+6) \sqrt{(k+1)(k+2)} \delta_{l (k+2)} + \nn \\
		&&+\,\, g\sqrt{(l+1)(l+2)(l+3)(l+4)} \delta_{k (l+4)} +\nn\\
		&&+\,\, g\sqrt{(k+1)(k+2)(k+3)(k+4)} \delta_{l (k+4)} \, .  \nn
\eeq
Due to the Kronecker delta functions most of the matrix elements 
are zero and the Hamiltonian matrix is band-diagonal with only five 
near-diagonals different from zero.

As the first step of RGP, we define the Hamiltonian with a big 
cutoff $N$, \ie, we reduce the space of states to the one spanned by 
the set $\{ |0\rangle, |1\rangle, \ldots, |N \rangle \}$ and write 
the Hamiltonian as $(N+1)\times (N+1)$ matrix with elements given 
by Eq.~(\ref{Hmn}). 

Then, we denote $\langle k | \psi \rangle = \psi_k$ and write the 
resulting Hamiltonian matrix eigenvalue equation in the form
\beq
\sum_{l=0}^{N} H_{kl} \psi_l \es E\psi_k.
\label{eigen}
\eeq
We extract $\psi_{N}$ from the highest energy equation, $k=N$, in the 
set (\ref{eigen}) of $N+1$ equations, and use it in the remaining $N$ 
equations in the set (\ref{eigen}) with $k<N$. This is essentially 
Gaussian elimination. The resulting set of only $N$ equations is
\beq
\sum_{l=0}^{N-1} \left( H_{kl} + 
	{ H_{kN} H_{Nl} \over E - H_{NN}} \right) \psi_l \es E \psi_k.
\label{eigenRG0}
\eeq
Taking into account that $H = H_0 + g H_I$, we arrive at the equation
\beq
\sum_{l=0}^{N-1} \bigg( H_{0\, kl} + \underbrace{
	g H_{I\, kl} + g^2{ H_{I\, kN} H_{I\, Nl} \over 	
	E - (H_{0\, NN} + g H_{I\, NN}) } }_{\displaystyle
	g H_{I\, kl}^{(1)}(E)} \bigg) \psi_l \es E \psi_k,
\label{eigenRG}
\eeq
which is an eigenvalue problem for a new matrix with elements $H_{0\, kl} + 
g H^{(1)}_{I\, kl}(E)$. We call it the matrix of effective Hamiltonian.
It can be used as a starting point for the next Gaussian elimination step,
but we notice that there appears a difficulty: $H_I^{(1)}$ depends on an 
unknown eigenvalue $E$. However, we are interested only in small 
eigenvalues of $H$, so small that they are negligible when compared with 
$H_{0\, NN} + g H_{I\, NN}$. The latter is the highest-index diagonal 
matrix element of $H$ that appears in the denominator on the left-hand 
side in Eq.~(\ref{eigenRG}). Thus, we bypass the difficulty by introducing 
the approximation 
\beq
E-(H_{0\, NN} + g H_{I\, NN}) &\approx& -(H_{0\, NN} + g H_{I\, NN}).
\label{E}
\eeq
This approximation limits our ability to precisely calculate eigenvalues $E$
using our RGP to lowest eigenvalues. For this price we can do as many Gaussian 
steps as we like, using the relation
\beq
H_{I\, kl}^{(j+1)} \es H^{(j)}_{I\, kl} - 
	g{ H^{(j)}_{I\, kn} H^{(j)}_{I\, nl} \over 
	H_{0\, nn} + g H^{(j)}_{I\, nn}} \, ,
\label{RG}
\eeq
with $n=N-j$, provided that we keep satisfying condition (\ref{E}) 
for every step number~$j$.
We may suppose that for some $M = N-j$, $H_{0\, MM} + g H^{(N-M)}_{I\, MM} 
\approx E$ and the condition (\ref{E}) may be eventually violated. On the 
other hand, we can continue a calculation assuming the simplification and 
using the approximate Gaussian steps to eliminate states $|N\rangle, 
|N-1\rangle, \ldots, |n+1\rangle$ 
down to a small $n$. We analyse results of such steps for a set of smallest 
eigenvalues for which the violation of our approximation appears quite small.

From now on, the Hamiltonian reduced in size by performing $N-n$ approximate 
Gaussian steps (in other words the renormalized Hamiltonian) is denoted by 
$H^{RG}_I(n)$. On the other hand, a Hamiltonian matrix reduced to the same 
size with a PC is denoted by $H^{PC}_I(n)$. This means that both $H^{RG}_I(n)$ 
and $H^{PC}_I(n)$ are $(n+1) \times (n+1)$ matrices but their matrix elements 
are different. 
The matrix elements of $H^{RG}_I(n)$ are to be calculated and the matrix 
elements of $H^{PC}_I(n)$ are directly given by the right-hand side of 
Eq.~(\ref{Hmn}) as the terms proportional to $g$.

Note that some matrix elements of the renormalized matrix depend on $N$ and 
$n$. If the dependence on $N$ were divergent or otherwise significant, it 
would have to be removed by counterterms in  $H^{(0)}$. Since in our model 
there are no divergences, and the limit of $N \rightarrow \infty$ is easily 
achieved for small eigenvalues, one can simply use some big $N$ and calculate 
the corresponding matrices of operators $H^{RG}(n)$. $N$ sufficiently large 
for working without counterterms to be valid, is determined in 
Sec.~\ref{comparison}. $H^{RG}(n)$ will have some matrix elements that 
vary with $n$. The only variation we study here is the one obtained assuming
condition~(\ref{E}). A more precise study than the one described here would 
be required to identify consequences of a finite ratio $E/H_{nn}$.

\section{ RGP evolution of Hamiltonian matrix elements} 
\label{sec:course}

\subsection{General band-diagonal Hamiltonian}

Matrix elements of interaction part of any band-diagonal Hamiltonian can be 
written in the form
\beq
H_{I\, kl} \es \sum_{i=-m}^m h_i(k) \delta_{k-l,i} \, 
\label{bdH}
\eeq
for some integer $m$. From Eq.~(\ref{RG}) follows that $H^{RG}_{I\,kl}(N-1) 
\neq H^{RG}_{I\,kl}(N) \equiv H^{PC}_{I\, kl}(N)$ only for such $k$ that 
$H^{RG}_{I\, kN}(N) \neq 0$ and $H^{RG}_{I\, Nl}(N) \neq 0$. This implies 
both $k,l \geq N-m$. Thus, after first
Gaussian step, done in appoximation (\ref{E}) or exactly, $H^{RG}_I$ is still
band-diagonal. Moreover, only matrix elements situated in $m\times m$ 
submatrix in highest-energy corner of $H^{RG}_I(N-1)$ can be different then
corresponding matrix elements of $H^{PC}_I(N-1)$. We see, that we can repeat
this reasoning recursively, with $N$ changed into $n$ (the highest index after 
some number of Gaussian steps). 

Thus, we can conclude that $H^{RG}(n)$ is band-diagonal for any $n$, and 
that not more then $m^2$ its matrix elements, located in the high-energy corner, 
does really depend on $n$. Eq.~(\ref{RG}) allows to write down the set of
(conjugated) recursions for these matrix elements. Numerical iteration of
such a set can be easily done using computer.

\subsection{Oscillator example}

First, we see that in the case of oscillator's Hamiltonian (\ref{Hmn}),
$m=4$. Further, from Eqs.~(\ref{Hmn}) and (\ref{RG}) we see, that no 
new nonzero matrix elements appear in the $4\times 4$ submatrix in the 
high-energy corner during the Gaussian step. Thus, only eight matrix 
elements depend on $n$. Finally, the symmetry of Hamiltonian matrix is 
preserved by RGP transformation, so there are only six independent 
functions of $n$ in $H_I^{RG}(n)$. To number them, we introduce a function 
\beq
i(k,l) \es \left\{
	\begin{array}{c @{{\quad \rm for} \quad} l}
	j	& k=l=n-(j\!-\!1) {\,\,\rm and\,\,} j \in \{1,2,3,4\} ,\\
	5	& (k,l) = (n,n\!-\!2) 	    {\,\,\rm or\,\,} (k,l) = (n\!-\!2,n),\\
	6	& (k,l) = (n\!-\!1,n\!-\!3) {\,\,\rm or\,\,} (k,l) = (n\!-\!3,n\!-\!1),
	\end{array} \right.
\label{ikl}
\eeq
and denote
\beq
\xi_{i(k,l)}(n) \es { H^{RG}_{I\, kl}(n) \over H^{PC}_{I\, kl} }.
\label{xi}
\eeq
This mean that $\xi_i(n\!\!=\!\!N) = 1$ and the deviation of $\xi_{i(k,l)}(n)$ 
from unity is the relative correction to $H^{PC}_{I\,kl}$ resulting from RGP. 
Values of $g\xi_i$ are effective coupling constants in the highest-energy 
$4 \times 4$ submatrix of $H_I^{RG}(n)$. 

From Eq.~(\ref{RG}), we obtain a set of recursions for coefficients $\xi_i(n)$. 
In the case of oscillator six conjugated first order recursive equations can 
be simply split into two sets of three equations of the second order. One such 
set is
\beq
\xi_1(n-2) \es \xi_3(n) 
	-g { f_{02} ^2  \over  f_{22}  } 
	{ \xi_5(n)^2 \over n + g f_{00} \xi_1(n) - E } 
	\label{RG1},\\
\xi_3(n-2) \es 1 
	- g { f_{04}^2 \over f_{44} } 
	{ 1 \over n + g f_{00} \xi_1(n) - E }
	\label{RG3},\\
\xi_5(n-2) \es 1 
	- g { f_{02} f_{04} \over f_{24} } 
	{ \xi_5(n) \over n + g f_{00} \xi_1(n) - E }
	\label{RG5},
\eeq
where $f_{kl} = H_{I\, (n-k)(n-l)}$ are functions of $n$, explicitly known 
from Eq.~(\ref{Hmn}). The second set is almost identical. The only differences
are that all down indices are increased by one and $(n-1)$ appears instead of 
$n$ in the denumerator on the right-hand side. For large $n$ the two sets 
evolve nearly identically. It is evident that the coefficients $\xi_i(n)$ 
are easy to calculate numerically.

\section{Comparison of RGP and PC}
\label{comparison}

In Eq.~(\ref{eigenRG}), we can see that $H_I^{PC}(n)$ can be interpreted 
as $H_I^{PC}(N)$ after $N-n$ Gaussian steps performed in the approximation
\beq
H_{I\, kl}  + g{ H_{I\, kN} H_{I\, Nl} \over E - H_{NN} } 
	&\approx& H_{I\, kl}.
\label{cut}
\eeq
This approximation neglects terms obtained in the approximation (\ref{E}).
Therefore, we can predict, that the results obtained using the RGP will be 
more accurate then the ones obtained using the PC. 

It is not clear how to compare the accuracies of the two procedures, since 
we don't know the exact eigenvalues of the infinite Hamiltonian. However, 
in our simple model, we can very precisely calculate small eigenvalues of 
the Hamiltonian numerically diagonalizing $H^{PC}(M)$ for some big cutoff 
$M$. 

In all computations, the numbers were handled with precision of 53 binary 
places (almost 16 decimal places). The numerical calculations show that 
for $M \in [200,1000]$ and $g \in \{0.01,0.1,1,10\}$ the three smallest 
eigenvalues, denoted $E_0$, $E_1$, $E_2$, do not depend on $M$ with a 
relative precision $10^{-10}$. For this reason, we can consider them to be 
accurate results and define the (relative) accuracy of RGP and PC as 
$|E^{RG}_i(n)-E_i|/E_i$ and $|E^{PC}_i(n)-E_i|/E_i$, respectively. Also 
for this reason $N = 200$ will be considered sufficiently large initial 
cutoff and it will be used as a starting point for RGP described in 
previous sections.

We shall illustrate our results on examples with three different values 
of the initial coupling constant: $g=0.01$, $g=1$ and $g=10$. These 
values are chosen because each one illustrates a different regime of RGP 
results. $g=0.01$ illustrates how RGP works in the case of interactions 
which only slightly change eigenvalues of $H_0$. $g=10$ illustrates how 
RGP works in the case of interactions dominating the system (which 
manifest itself in eigenvalues much different then those of $H_0$). 
$g=1$ illustrates the intermediate situation. The results are discussed 
separately for different values of $g$.

\subsection{ Results for $g=0.01$ }

For $g=0.01$, the accurate eigenvalues are $E_0 = 0.1687726041$, 
$E_1 = 1.716925770$ and $E_2 = 3.602838696$. Both procedures, RGP 
and PC, give precise results. Even for $n$ as small as $n=10$, 
$E_0$ is recovered with accuracy 
$4\cdot 10^{-10}$ and $E_1$ and $E_2$ are recovered with accuracy 
$2 \cdot 10^{-8}$ in the case of RGP. In the case of PC, the 
accuracy of these three eigenvalues is about $10^{-7}$. 
We see that RGP increases accuracy of predicting $E_0$ about 500 
times. Also $E_1^{RG}$ and $E_2^{RG}$ are calculated about 5 times
more precisely then $E_1^{PC}$ and $E_2^{PC}$. Nevertheless, accuracy
obtained with PC is also brilliant.

\subsection{ Results for $g=1$ }

For $g=1$ and $n=10$, eigenvalues $E_1^{PC}$ and $E_2^{PC}$ differ from 
eigenvalues $E_1^{RG}$ and $E_2^{RG}$ much more then for $g=0.01$. We 
can say that the difference is qualitative. The correct eigenvalues are 
$E_1 = 3.521565666$ and $E_2 = 7.263980184$. $E^{RG}_2$ has accuracy 3\%, 
while $E^{PC}_2$ only 21\%. For $E_1$, the difference is 0.25\% to 6\% 
in favor of RGP. 

The correct ground state energy is $E_0 = 0.6487889141$. The errors are 
$2\cdot 10^{-4}$ in RGP and $6 \cdot 10^{-3}$ in PC. Both values can be 
considered satisfactory if the required accuracy is of the order of 1\%.

\subsection{ Results for $g=10$ }

Utility of RGP is most clearly visible in the case of $g=10$.
This is giant, but still realistic coupling. For example the pion-nucleon 
coupling constant is about 13~\cite{alpha_pn}. Moreover, there is no 
{\it a priori} reason for values of couplings emerging from SRG procedure 
to be small.

\begin{figure}[ht]
\centering
\includegraphics[width=.75\textwidth]{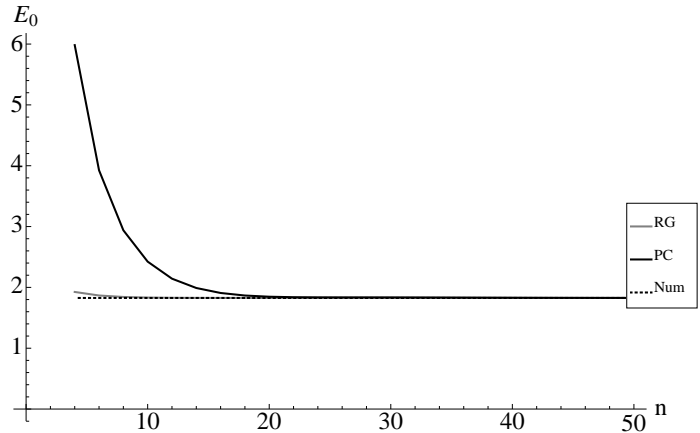} 
\caption{The ground state energy of $H$ for $g=10$ 
	 determined with RGP and PC, compared with the value achieved 
	 in the accurate numerical calculation, Num. }
\label{E0}
\end{figure}

The accurate eigenvalues in this case are $E_0 = 1.826275924$, 
$E_1 = 7.790412053$ and $E_3 = 15.70695963$. The results for eigenvalue $E_0$, 
obtained with RGP and with PC, as a function of cutoff $n$
are plotted in Fig.~\ref{E0}. One sees that PC with highest index
smaller than $\sim$15 produces a huge error in the eigenvalue. 
In the same cutoff region, RGP still gives a reasonable value of $E_0$. 

For $n=10$, $E_0^{RG}$ has accuracy 0.35\%, which is about 
100 times better than 33\% obtained with PC. These results 
prove that as long as $E \ll H_{nn} \sim n^2$, RGP gives a 
reasonable estimation of an eigenvalue. Even for $n=4$, RGP 
reproduces $E_0$ with 5.5\% accuracy while PC introduces 
about 230\% of error.

For $n=4$, the eigenvalues $E_1$ and $E_2$ are obtained 
with poor accuracy. In RGP, $E_1^{RG}$ and $E_2^{RG}$ have
errors 11\% and 43\%, respectively. Obtaining $E_1^{PC}$ and 
$E_2^{PC}$ does not make any sense.

\section{Effective Hamiltonians}
\label{EH}

The important question is how it happens that the RGP 
produces a better result then PC. In PC, there appears 
a non-physical parameter $n$. If $n$ is not large
enough, even small eigenvalues depend on $n$. The key 
idea behind the RGP is to base physical predictions 
(in this case the small eigenvalues of $H$) on the 
effective theory whose Hamiltonian $H_I^{RG}(n)$ is
calculated from the initial one, instead of plainly 
cutting off the initial theory to the arbitrary number 
of states.

As it was explained in Sec.~\ref{sec:course}, in the case of 
band-diagonal Hamiltonian most of the matrix elements of the 
effective Hamiltonian stay invariant under the RGP transformation 
given by Eq.~(\ref{RG}).

\begin{figure}[ht]
\centering
\includegraphics[width=.75\textwidth]{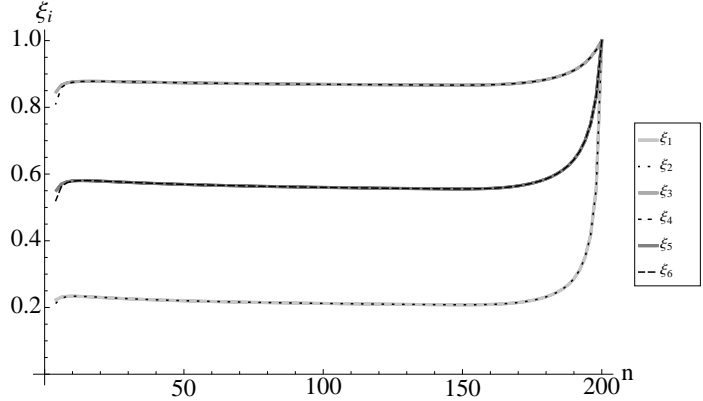}
\caption{The ratios of renormalized matrix elements to the 
	 original ones, $\xi_i$, as functions of $n$, obtained for the initial 
	 cutoff $N=200$ and $g = 10$, see Eq.~(\ref{xi}).}
\label{xi-fig}
\end{figure}

The ratios of the changed terms in $H_I^{RG}(n)$ to the corresponding 
terms in $H^{PC}_I$ for $g=10$ are presented in Fig.~\ref{xi-fig}. One can 
see that for $n < N$
\beq
0< \xi_1 \approx \xi_2 < \xi_5 \approx \xi_6 < \xi_3 \approx \xi_4 < 1 \, .
\eeq
This means that these renormalized matrix elements are 
smaller then the original ones, but the sign remains unchanged. 
The approximate equalities correspond to the observation made 
in Sec.~\ref{sec:course} that the six evolving matrix elements 
form two similar independent subsets.

Within the $4 \times 4$ highest-energy corner of the 
Hamiltonian matrix, the closer is the matrix element 
to the cutoff corner, the bigger is the necessary 
correction. It is worth noticing that although $\xi_i$ 
considerably depart from unity, they still very weakly 
depend on $n$. Calculations for various $g$ show that 
they also weakly depend on $g$.

\section{Conclusion}
\label{Conclusion}

We have considered a simple version of RGP in application to an elementary 
Hamiltonian of a single-mode scalar field with an anharmonic interaction 
term proportional to $\varphi^4$. We have calculated the ground state and 
two least excited states energies and compared results obtained using RGP 
to the results obtained using PC.

Our results show that when the coupling constant $g$ is large, the RGP is 
much more accurate than the PC procedure. For $g=10$, RGP reproduces $E_0$ 
with 5.5\% accuracy in a $5\times 5$ matrix, which can be considered a 
reasonable estimate. Such small matrices provide a model of small spaces 
of states that one can handle non-perturbatively using computers in realistic 
theories. In the matrix of the same size, PC gives the result with an enormous 
error (about 230\%). RGP is significantly more accurate also for small values 
of $g$.

In the presented oscillator model, the improvement due to RGP is \linebreak
achieved through a calculation of eight cutoff-dependent terms (only six of them 
being different) in the effective Hamiltonians with small cutoffs. A small number 
of terms is an unavoidable consequence of the band-diagonal structure of 
the initial Hamiltonian. This feature make RGP a potentially convenient tool
for analysing band-diagonal Hamiltonians, in particular these emerging 
form applying SRG to some realistic field-theoretic Hamiltonians.

The author would like to thank Stanis{\l}aw G{\l}azek for many discussions.

\end{document}